\documentclass{SCIS2019}

\begin{document}
\ArticleType{REVIEW}
\Year{2019}
\Month{}
\Vol{}
\No{}
\DOI{}
\ArtNo{}
\ReceiveDate{}
\ReviseDate{}
\AcceptDate{}
\OnlineDate{}

\title{Pushing AI to wireless network edge: An overview on   integrated sensing, communication, and computation towards 6G}{Pushing AI to wireless network edge: An overview on   integrated sensing, communication, and computation towards 6G}

\author[1]{Guangxu Zhu}{}
\author[2]{Zhonghao Lyu}{}
\author[1,3]{Xiang Jiao}{}
\author[1,3]{Peixi Liu}{}
\author[4]{Mingzhe Chen}{}
\author[2]{\\Jie Xu}{{xujie@cuhk.edu.cn}} 
\author[1,2,6]{Shuguang Cui}{{shuguangcui@cuhk.edu.cn}}
\author[5,6]{Ping Zhang}{}

\AuthorMark{Zhu G}

\AuthorCitation{Zhu G, Lyu Z, Jiao X, et al}


\address[1]{Shenzhen Research Institute of Big Data, Shenzhen {\rm 518172}, China}
\address[2]{Future Network of Intelligence Institute (FNii) and School of Science and Engineering (SSE),\\ The Chinese University of Hong Kong (Shenzhen), Shenzhen {\rm 518172}, China}
\address[3]{State Key Laboratory of Advanced Optical Communication Systems and Networks, School of Electronics,\\
Peking University, Beijing {\rm 100871}, China}
\address[4]{Department of Electrical and Computer Engineering
and Institute for Data Science and Computing,\\ University of Miami, Coral Gables {\rm 33146}, USA}
\address[5]{State Key Laboratory of
Networking and Switching Technology,\\ Beijing University of Posts and Telecommunications, Beijing {\rm 100876}, China}
\address[6]{Peng Cheng Laboratory, Shenzhen {\rm 518066}, China}

\abstract{Pushing artificial intelligence (AI) from central cloud to network edge has reached board consensus in both industry and academia for materializing the vision of artificial intelligence of things (AIoT) in the sixth-generation (6G) era. This gives rise to an emerging research area known as edge intelligence, which concerns the distillation of human-like intelligence from the huge amount of data scattered at wireless network edge. In general, realizing edge intelligence corresponds to the process of sensing, communication, and computation, which are coupled ingredients for data generation, exchanging, and processing, respectively. However, conventional wireless networks design the sensing, communication, and computation separately in a task-agnostic manner, which encounters difficulties in accommodating the stringent demands of ultra-low latency, ultra-high reliability, and high capacity in emerging AI applications such as auto-driving. This thus prompts a new design paradigm of seamless integrated sensing, communication, and computation (ISCC) in a task-oriented manner, which  comprehensively accounts for the use of the data in the downstream AI applications. In view of its growing interest, this article provides a timely overview of ISCC for edge intelligence by introducing its basic concept, design challenges, and enabling techniques, surveying the state-of-the-art development, and shedding light on the road ahead.}


\maketitle

\section{Introduction}

With the commercialization of the fifth-generation (5G) wireless networks, we are moving towards a new  era with everything connected. The convergence of modern information and communications technology (ICT) technologies such as the internet of things (IoT), cloud computing, mobile edge computing (MEC), and big data analytics is prompting a huge leap in social productivity and management efficiency. At the same time, artificial intelligence (AI) witnesses great success in various applications and continues its explosive growth and penetration to  all walks of life. The great success achieved by AI has driven the ongoing convergence of the communication networks with the AI technology. Particularly, in the current 5G network, AI has been used as an add-on module to boost the network performance, while in the future sixth-generation (6G) era, AI will be deeply integrated in the network design to achieve the so-called AI-native network. As a consequence, the future 6G wireless communication networks will go beyond a pure data delivery pipeline, and become a comprehensive platform integrating sensing, communication, computation, and intelligence to deliver pervasive AI services \cite{You2021SCIC-6G,Letaief2022JSAC-edge-overview,feng2021joint}.

\subsection{Edge AI}

The interplay between wireless communication networks and AI technology can be generally classified into two categories, namely the AI-assisted communication\cite{Letaief2019CM-6G} and communication-assisted AI\cite{Shen2022Intelligence-survey}. Specifically, AI-assisted communication, which refers to the use of AI to improve existing communication systems (e.g., physical layer modules such as channel estimation and signal detection \cite{Ye2018WCL-AI-estimation}), boosts the end-to-end performance of communication links to achieve higher rates, lower latency, and wider connectivity. On the other hand, communication-assisted AI, which refers to the use of communication networks to help AI acquisition, allows for distributed AI training and inference across the entire network, as well as pervasive AI services delivery. 

The focus of the current work is on the communication-assisted AI, which has received increasing attention from academia and industry in recent years, as various AI applications, e.g., industrial Internet, smart cities, smart health, and auto-driving mature and gain popularity.  Traditional communication-assisted AI based on cloud data processing, which requires the delivery of a large amount of data collected by the edge devices to the cloud for AI distillation, cannot support emerging mission-critical AI applications such as industrial control, virtual reality (VR), augment reality (AR), and auto-driving due to the following challenges.

\begin{figure*}[t]
	\centering
	\subfloat[Number of mobile and IoT connections \cite{Cisco-annual}.]{\includegraphics[width=0.485\textwidth]{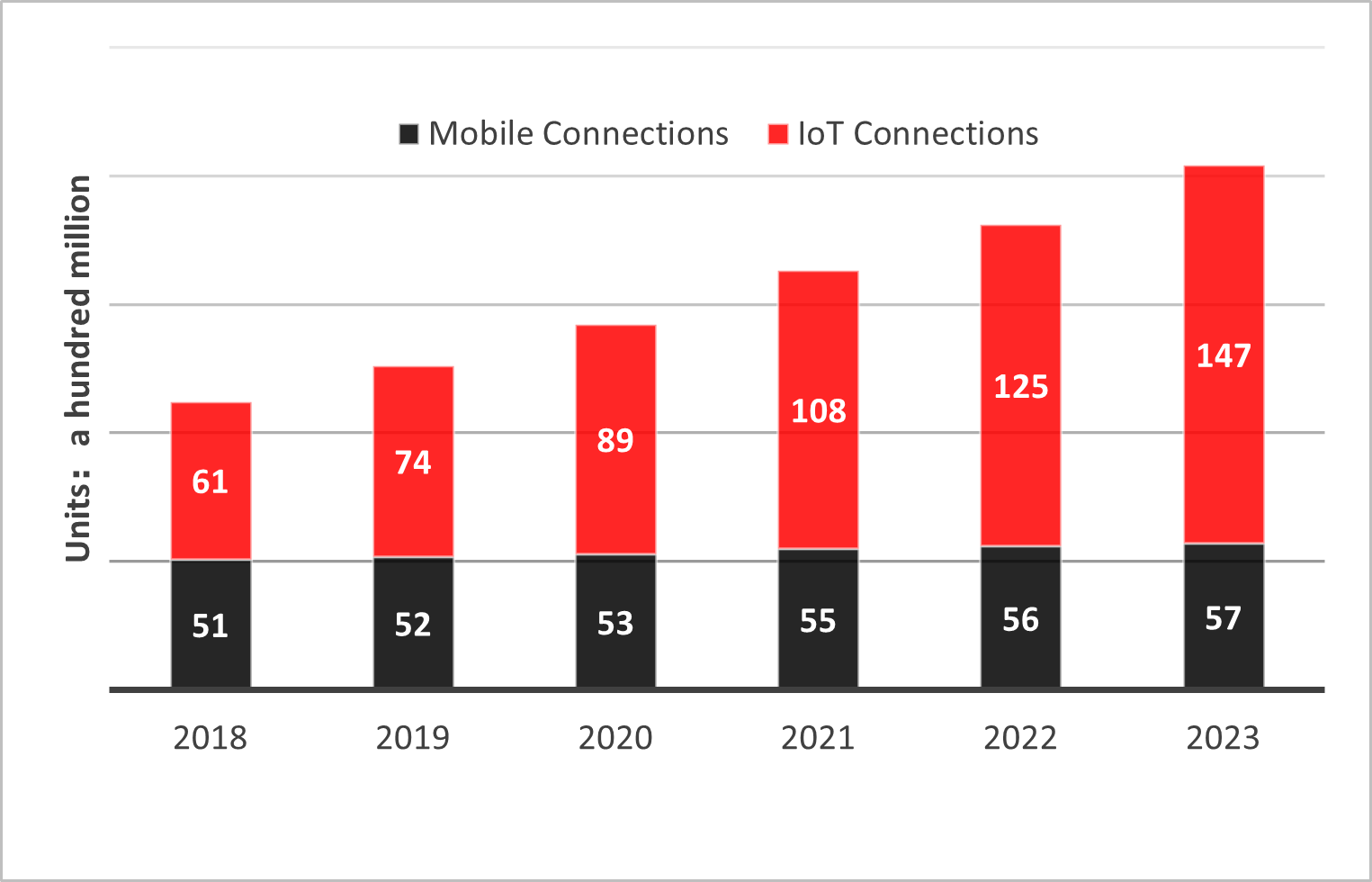}
		\label{fig:number of connections between mobile and IoT}}\hfil
	\subfloat[Global data traffic growth \cite{Cisco-global}.]{\includegraphics[width=0.485\textwidth]{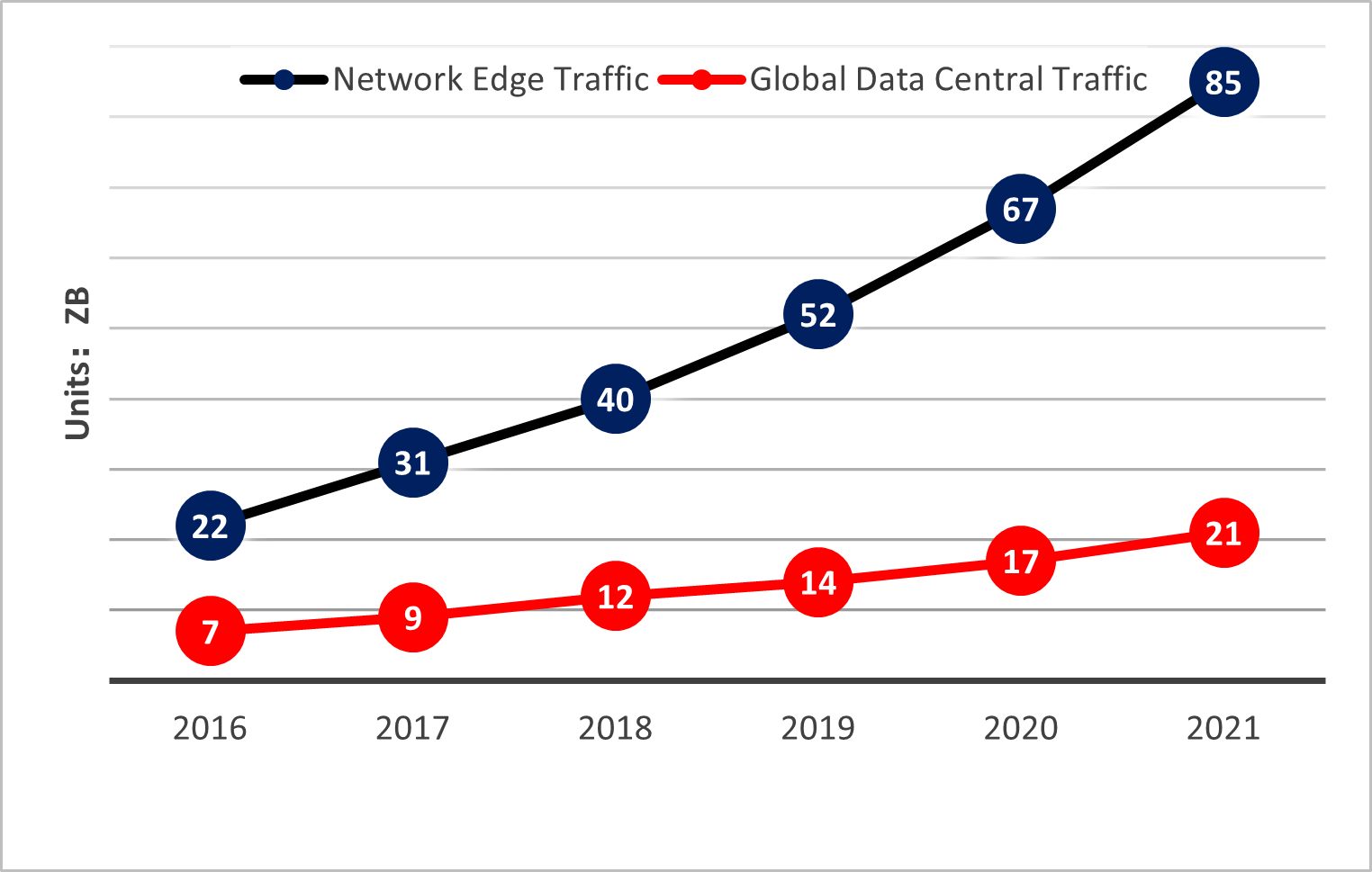}
		\label{fig:global traffic growth chart}}
	\caption{Edge AI connections and data growth.}
	\label{fig:Edge AI Data Chart}
\end{figure*}

\begin{itemize}

   \item{\bf Massive data and network connectivity}: According to Cisco, global business volumes increased at a rate of nearly 42$\%$ per year between 2018 and 2020. As shown in Fig.  \ref{fig:number of connections between mobile and IoT}, the total number of connected devices worldwide is expected to reach 29.3 billion by 2023, with 5.7 billion mobile connections and 14.7 billion IoT connections \cite{Cisco-annual}. According to Huawei \cite{Huawei-2030}, the total number of global network connections will reach 200 billion by 2030, with wireless and passive connections accounting for roughly half of the total. Aside from the massive number of temperature, humidity, pressure, and photoelectric sensors in the industrial sector, the network will also include a large number of smart vehicles, robots, and drones. The constant generation of large amounts of data and network connections increases the demand for communication capacity and computing power.

   \item{\bf Data sinking}: Previously, big data, such as online shopping records, social media content, and business information, was primarily generated and stored in hyperscale data centers. However, with the proliferation of mobile and IoT devices, this trend is now being reversed. Cisco predicts that by 2021, all people, machines, and things will generate nearly 85 Zettabytes of usable data at the network edge. In sharp contrast, as shown in Fig. \ref{fig:global traffic growth chart}, global data center traffic will reach only 21 Zettabytes by 2021 \cite{Cisco-global}. In the traditional cloud computing design, the massive data sinking to the network edge need to be transferred to a central cloud server away from the edge for analysis and processing. This would undoubtedly result in unacceptable communication cost and delay.

   \item{\bf Ultra-low latency requirements}: The majority of new AI services require high-demand network connections with ultra-low latency and ultra-high reliability. For example, smart industrial Internet  requires real-time state feedback, data analysis, and highly-accurate control \cite{You2021SCIC-6G}. Similarly, VR and AR applications request real-time aggregation, analysis, and reconstruction on three-dimensional (3D) images for complex control feedback. In these use cases, the required closed-loop sensing-communication-computation latency must be within around one millisecond, which poses great challenges to the current communication networks.
\end{itemize}

To address these issues, industry and academia have reached consensus on bringing computing power and AI functionality close to data, known as edge computing or fog computing \cite{Shi2016IoTJ-edge}. As its name implies, edge computing aims to relocate some of the data processing and data storage for specific services from the central cloud to the distributed edge network nodes, which are more physically and logically close to the data provider, so as to achieve the desired low latency. AI, on the other hand, seeks to emulate human intelligence in a machine by learning from the data. Naturally, the convergence of edge computing and AI gives rise to a new area known as edge AI \cite{Cisco-global}, which aims to provide mobile terminals with low-latency AI services by exploiting both the computing resources and data scattered at the network edge. In view of its promising performance gain, edge AI has received increasing attention from both academia and industry, which is becoming a hot research area in the direction of communication-assisted AI \cite{Letaief2022JSAC-edge-overview,Zhou2019-edge-overview,Park2019Proceed_edge}.


\begin{figure}[t]
	\centering
	\includegraphics[width=\textwidth]{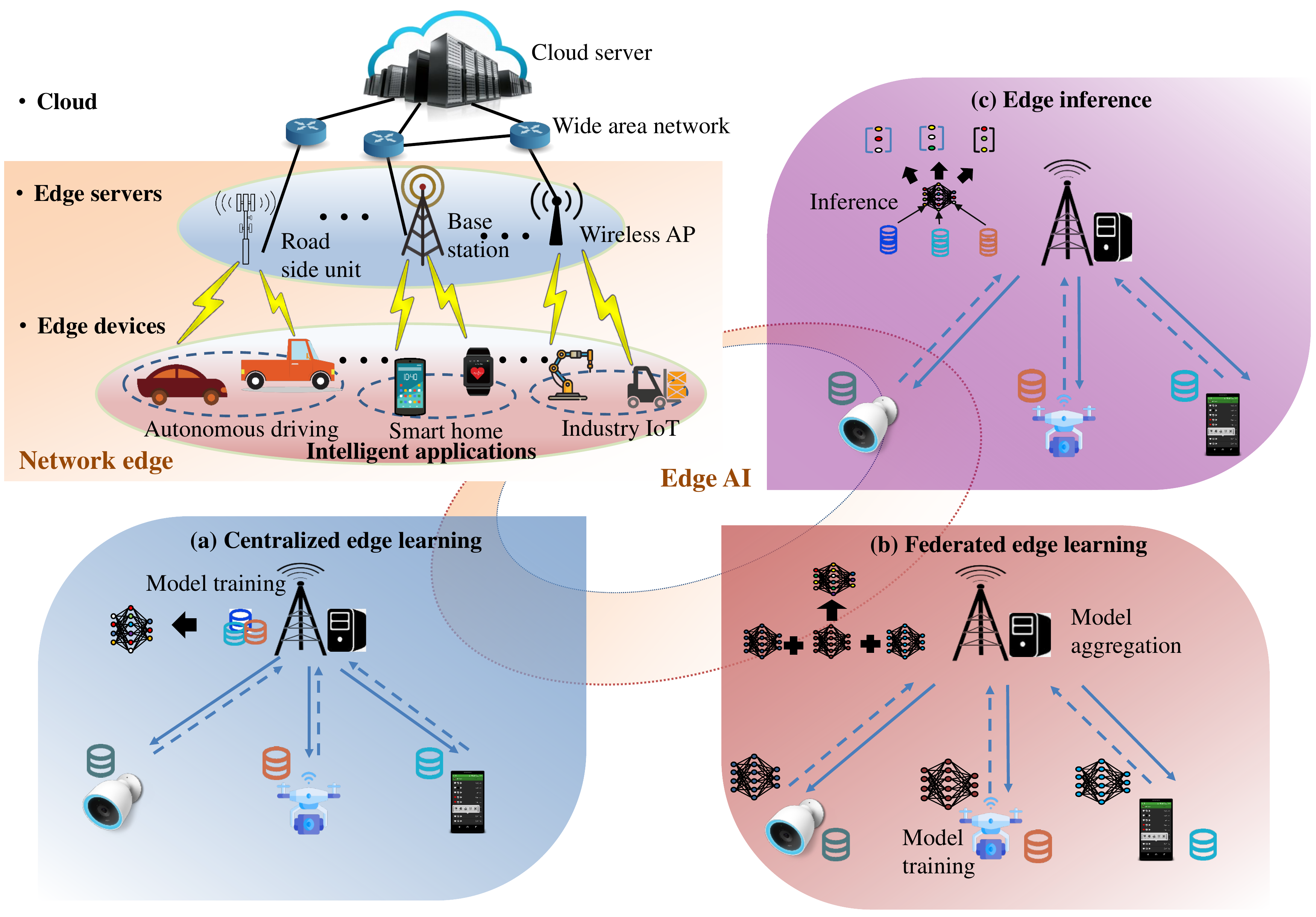}
	\caption{The concept of edge AI and three related scenarios.}
	\label{fig:edge-AI}
\end{figure}

 As shown in Figure \ref{fig:edge-AI}, learning in edge AI can be divided into two types, i.e., centralized edge learning and distributed edge learning, depending on where the data is processed. After the AI model is well-trained, it can be deployed for edge inference.
 In the early stage of the development of edge AI, AI model is attained via centralized edge learning in many large AI companies (e.g., Google, Facebook, and Microsoft) \cite{Zhou2019-edge-overview}. However, centralized edge learning requires uploading the private raw data (e.g., personal photos at the smart phones)  from edge devices to edge server, which may pose grand challenge to the user data privacy.  
 On the other hand, thanks to the well-known Moore's law, the power of computing chips such as central processing unit (CPU) and graphics processing unit (GPU) has been continuously upgraded with decreasing cost. In particular, with the emergence of dedicated AI chipsets, the computing power of edge devices become more and more powerful and can support the running of machine learning (ML) tasks. This thus drives the rapid development of distributed edge learning, such as federated edge learning (FEEL), to exploit the rich distributed computing resources at the network edge. Moreover, an additional bonus of the FEEL is the data privacy preservation due to the waiving of the need to upload the raw data, but sharing the less privacy-sensitive gradient or model updates.

\subsection{Integrated Sensing, Communication, and Computation (ISCC)}

\begin{figure}[t]
	\centering
\includegraphics[width=\textwidth]{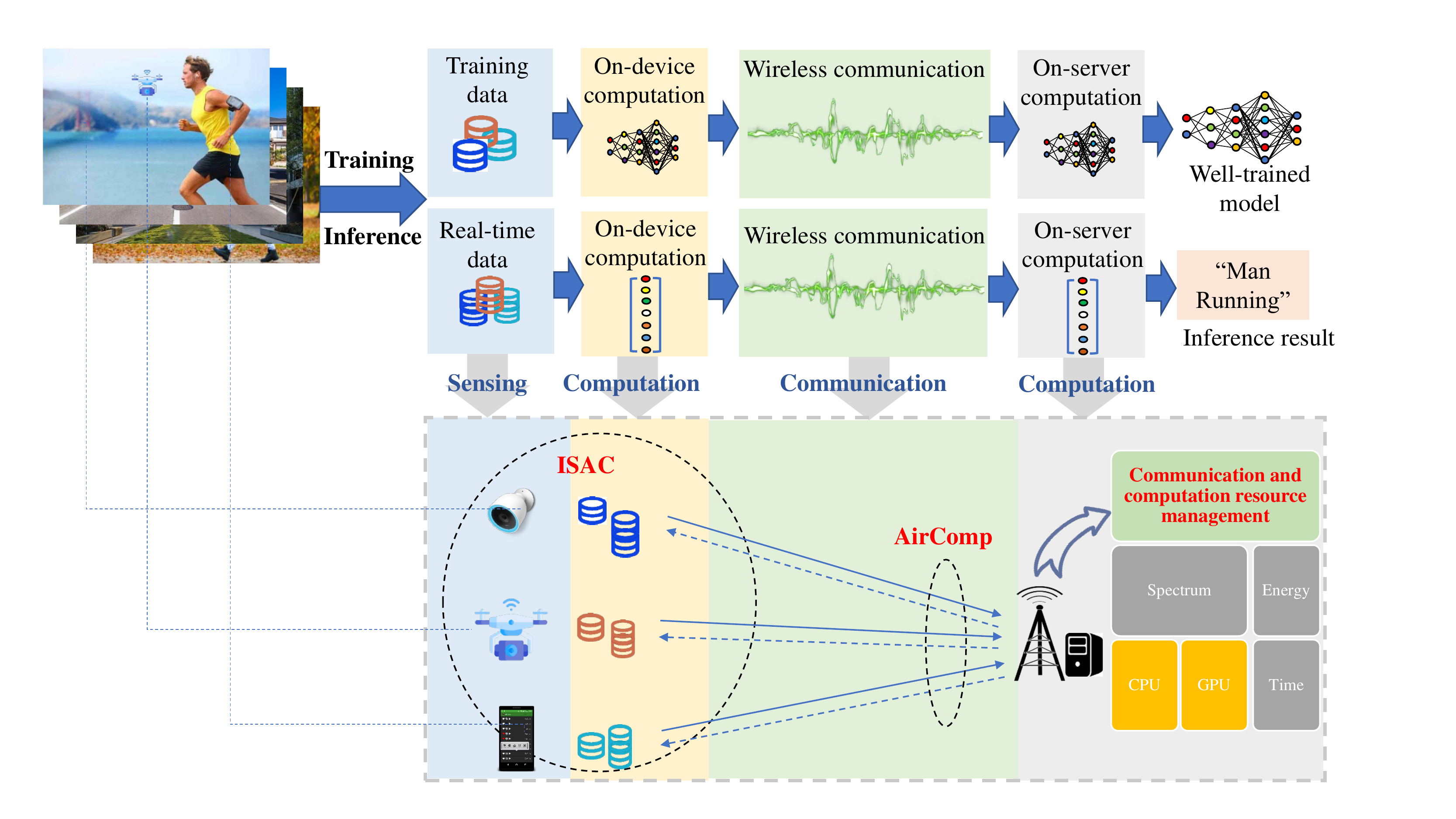}
	\caption{Sensing, communication, and computation in edge AI.}
	\label{fig:ISCC}
\end{figure}

In practice, a complex system (e.g., the edge AI system) generally consists of three coupled processes, namely sensing, communication, and computation, as shown in Figure \ref{fig:ISCC}. However, in traditional wireless networks, these three processes are designed separately for different goals: sensing for obtaining high-quality environmental data, communication for data delivery, and computation for executing the downstream task within a certain deadline. Such a separation design principle encounters difficulty in accommodating the stringent demands of ultra-low latency, ulta-high reliability, and high capacity in emerging 6G applications such as auto-driving. This thus prompts a new wireless design paradigm of integrated sensing, communication, and computation (ISCC) in a task-oriented manner, which comprehensively accounts for the use of the data in the downstream tasks (e.g., AI applications) in 6G.
In the literature, there have been some prior studies on the integration of two of the above three entities. Some examples include joint communication and computation resource management, over-the-air computation (AirComp), and integrated sensing and communication (ISAC), which are detailed in the following.


\subsubsection{Joint Communication and Computation Resource Management}
Data acquisition and model computation are typically separate processes in traditional complex communication systems. With the rapid increase in data volumes in MEC, the communication and computation capabilities at the network edge become the bottleneck. Particularly, the limited wireless resources make it challenging for the edge server to receive significant amounts of data from edge devices swiftly through wireless links. 
Hence, many researches have focused on joint communication and computation resource management to tackle this issue in MEC. For example, in \cite{Chen2016ICC}, in order to minimize the energy and delay cost of the multi-user multi-task MEC system, the authors used a separable semidefinite relaxation method to jointly optimize the offloading decision and communication resource allocation. In \cite{Hoang2012WCNC}, in order to solve the resource allocation problem for MEC, the authors proposed an effective solution to maximize the quality of service (QoS) of all mobile devices, by transforming the problem into a linear programming model. Besides, in \cite{Mao2016GLOBECOM}, by using an online algorithm based on Lyapunov optimization, the energy-latency trade-off problem of multi-user MEC systems is studied, where the computation tasks arrive at the mobile devices in a stochastic manner.  
In addition, in \cite{Wang2018TWC}, the authors solved the problem of reducing the total energy consumption at the edge server in wireless-powered multi-user MEC systems, by jointly optimizing the AP's energy transmission beamforming, the user's central processing unit frequency and number of offloaded bits, and the time allocation among users to improve MEC performance.
Furthermore, joint communication and computation cooperation was exploited in MEC in \cite{Cao2019ITJ}, where a neighboring user node was enabled as not only a relay node for helping task offloading of a user, but also a computation helper to help remotely execute some of tasks of that user.
Joint communication and computation resource management can theoretically ensure the information security, mobile energy saving, and so on.

\subsubsection{AirComp}
 AirComp has emerged as a promising technology in recent years. AirComp, as opposed to ``communication before computing'', integrates computing into communication, resulting in a new scheme of ``communication while computing''. In contrast to traditional wireless communication over a multi-access channels (MAC), which requires separate transmission and decoding of information, AirComp allows edge devices to simultaneously transmit their respective signals on the same frequency band with proper processing, such that the functional computation of the distributed data is accomplished directly over the air. This thus significantly improves the communication and computing efficiency, and considerably reduces the latency required for multiple access and data fusion.

In \cite{Zhu2021WC-aircomp}, the authors provide an comprehensive overview of AirComp by introducing the basic principles, discussing the advanced techniques and several promising applications, and identifying promising research opportunities.
In order to achieve reliable AirComp in practice, the authors in \cite{Cao2021arXiv_power} focused on the power control problem in AirComp, and the optimal power allocation under both deterministic and fading channels were derived by minimizing the mean-squared error (MSE) of the aggregated signals. 
Similarly, the authors of \cite{Liu2020TWC} minimized the computation MSE at the receiver by optimizing the transmitting and receiving policy under the maximum power constraint of each sensor.
While only single cell was considered in \cite{Cao2021arXiv_power} and \cite{Liu2020TWC}, the power control problem of AirComp in the multi-cell scenario was considered in \cite{Cao2020TWC-multi-cell-aircomp}. To quantify the fundamental AirComp performance trade-off among different cells, in \cite{Cao2020TWC-multi-cell-aircomp}, the Pareto boundary of the multi-cell MSE region was characterized by minimizing the MSE subject to a set of constraints on individual MSE.
Note that the work in \cite{Cao2021arXiv_power,Liu2020TWC,Cao2020TWC-multi-cell-aircomp} only considered the scenario with single antenna. 
The authors of \cite{Zhu2019IoT} intended to develop multiple-input-multiple-output (MIMO) AirComp in order to achieve multi-modal sensing with high mobility, and design MIMO-AirComp equalization and channel feedback techniques for spatially multiplexing multi-function computation. 
Besides, some complicated systems with AirComp have been also considered in the literatures. For example, the authors of \cite{Feng2021TC} considered to use reconfigurable intelligent surface (RIS) to assist AirComp. Besides, in \cite{Zhang2022COMML}, under imperfect channel state information, the authors investigated the joint optimization of transceiver and RIS phase design for an AirComp system. In \cite{Fu2022TWC}, when the ground receiver is unavailable, unmanned aerial vehicles (UAVs) are utilized to establish line-of-sight (LoS) connections by tracking mobile sensors, and thus improving the performance of AirComp.

\subsubsection{ISAC}

ISAC generally refers to the integration of sensing and communication into a unified design in wireless networks to enhance the utilization efficiency of scarce spectrum and wireless infrastructures, pursue a mutual benefit via sensing-assisted communication and communication-assisted sensing \cite{Liu2021ISAC-survey}. In comparison to traditional wireless networks, ISAC can use the wireless infrastructure as well as limited spectrum and power resources for both communication and sensing, which could improve the  system performance at a lower cost.

ISAC is one of the potential key technologies in 6G networks that has received a lot of attention in the literature. 
Many works have focused on joint sensing and communication in \cite{Liu2020TC-JRC-overview}.
For example, in \cite{Liu2018TSP}, the authors proposed a dual-functional MIMO radar communication system that consists of a transmitter with multiple antennas that can communicate with downlink cellular users and detect radar targets at the same time. 
In \cite{Liu2020JRC-MIMO}, a joint transmit beamforming model for a dual-function MIMO radar and multiuser MIMO communication transmitter was proposed and the weighting coefficients of the radar beamforming were designed.
In \cite{Hua2021ISAC-beamforming}, the authors also considers the beamforming optimization problem in ISAC system, and the radar sensing performance is maximized subject to the communication users' minimum signal-to-interference-plus-noise ratio (SINR) requirements and the transmit power constraint of the base station (BS). 
The authors in \cite{Liu2022TSP-CR} employ the Cramer-Rao bound (CRB) as a performance metric of target estimation, and the CRB of radar sensing is minimized while guaranteeing a pre-defined level of SINR for each communication user.
In \cite{ZLyu2021}, the authors considered a UAV-enabled ISAC system, where UAV trajectory/deployment and beamforming design are jointly considered to balance the sensing-communication performance trade-off under quasi-stationary and mobile UAV scenarios, respectively. 
Furthermore, because of the benefits of spectrum sharing, ISAC has been used in a variety of systems. Furthermore, using RIS to facilitate radar sensing and ISAC has attracted growing research interests (see, e.g., \cite{Song2022ISAC_RIS,Song2022ISAC_RIS_WCNC,Wang2022TVT,Shi2022WC,Li2022I6GNet}). For instance, the authors in \cite{Song2022ISAC_RIS} derived the fundamental CRB for RIS-enabled NLOS sensing, and those in \cite{Song2022ISAC_RIS_WCNC} jointly designed the transmit beamforming at the transmitter and the reflective beamforming at the RIS for ensuring both sensing and communication performances. The authors of \cite{Wang2022TVT} used RIS for a joint design of constant-modulus waveform and discrete phase shift to mitigate multi-user interference in ISAC. In addition, in the overview paper \cite{Shi2022WC}, the authors elaborated the benefits of RIS in wireless communication, sensing, and security, and envisioned that the RIS-assisted communication and sensing would mutually benefit each other. The authors of \cite{Li2022I6GNet} considered the combination of ISAC and AirComp to improve the spectral efficiency and sensing performance, and the beamformers for sensing, communication, and computation were jointly optimized. 
The authors of \cite{Huang2020JDSCSSH} used ISAC in smart homes to provide inconspicuous sensing and ubiquitous connectivity. 
Besides joint sensing and communication, there are also a line of research on sensing-assisted communication. For example, a radar-assisted predictive beamforming design for vehicle-to-infrastructure communication was investigated in \cite{Liu2020TWC-Vehicular}, and it was found that the communication beam tracking overheads can be drastically reduced by exploiting the radar functionality of the road side unit.

\subsubsection{Task-oriented ISCC Towards Edge AI}
As previously stated, sensing, computation, and communication have a symbiotic relationship, especially in the context of edge AI. Specifically, the ultimate performance of edge AI depends on the input feature vector’s distortion level arising from three processes, i.e., data acquisition (sensing), feature extraction (computation), and feature uploading to edge server (communication). Particularly, sensing
and communication compete for radio resources, and the allowed communication resource further determines the required quantization (distortion) level such that the quantized features can be transmitted reliably to the edge server under a delay constraint. Thereby the three processes are highly coupled and need to be jointly considered. Furthermore, the implementation of  ISCC should be designed under a new task-oriented principle that concerns the successful completion of the subsequent AI task. Different from conventional communication system design aiming at maximizing the data-rate throughput, the ultimate  performance metrics of interest for the system become the inference/training accuracy, latency, and energy efficiency, etc. For instance, an edge AI task-oriented ISCC scheme can be designed to maximize the  inference/training accuracy under constraints on low latency and on-device resources. This is in sharp contrast to the classic separation-based design approach that consider the sensing, communication and computation processes in isolation. More examples of the task-oriented ISCC will be discussed in the subsequent sections.



\subsection{Structure of the Survey}

\begin{figure}[t]
	\centering
	\includegraphics[width=0.9\textwidth]{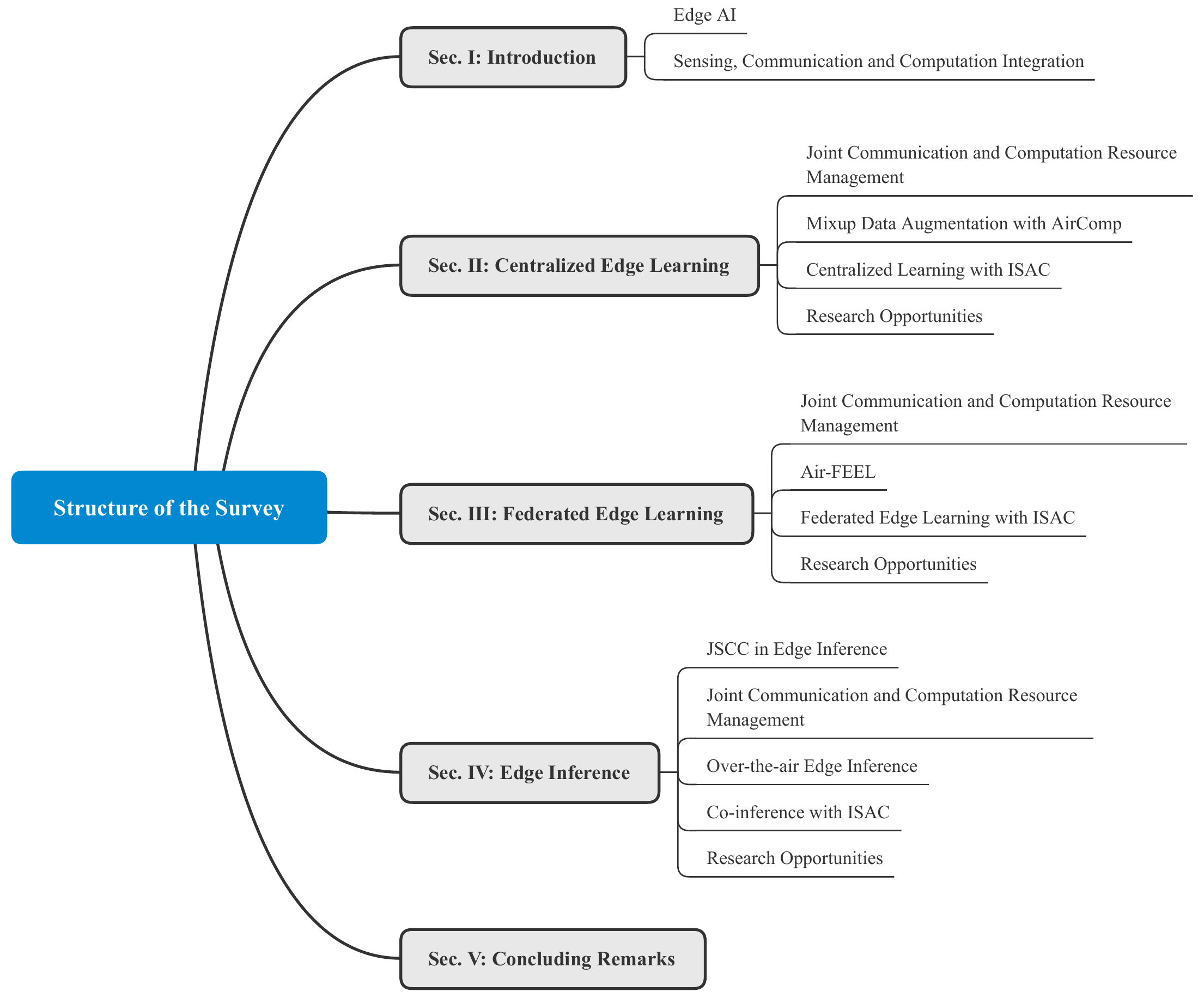}
	\caption{The structure of this article.}
	\label{fig:structure}
\end{figure}

Different from prior works considering the integration of sensing and communication or computation and communication in generic wireless networks, we focus on their integration towards edge AI applications. As a result, we classify them based on three application scenarios, i.e., centralized edge learning, federated edge learning, and edge inference. The remaining of the survey is organized as shown in Figure \ref{fig:structure}. 
In Section \ref{sec:centralized}, we first review the joint communication and computation resource management, mixup data augmentation with AirComp, and ISAC, respectively, in centralized edge learning, and then discuss the related research opportunities. 
In Section \ref{sec:FEEL}, we discuss the joint communication and computation resource management in federated learning, the application of AirComp into FEEL, and the combination of ISAC and FEEL, as well as the related research opportunities. 
In Section \ref{sec:inference}, we first discuss the joint source and channel coding, then present joint communication and computation resource management in edge inference, over-the-air edge inference, and co-inference with ISAC, respectively, and highlight several key research opportunities.
Finally, Section \ref{sec:conclusion} concludes the article. 


\section{Centralized Edge Learning}\label{sec:centralized}

With the continuing development of deep learning (DL) in recent years, the increasing DL model  complexity has posed a grand challenge in training due to the demand for computation power and storage capacity. There are two basic strategies to accommodate the increasing demands for the resources required by DL at the centralized cloud: scaling-up, which involves adding extra processing and storage resources to a single central server, and scaling-out \cite{Xu2022arXivSemantic}, which involves forming a server cluster by networking multiple servers each with certain computing and storage capacity. Recent rapid development of MEC makes it possible to deploy the mentioned two strategies at the network edge so as to ``bring the computation power close to the data'', leading to an emerging research area known as centralized edge learning.  

In the centralized edge learning, the data is first collected on the client side and then transferred through the wireless channel to a central edge server. The central server then stores and processes the data, and finally returns the learnt model back to the client. This architecture is simple to deploy and manage, particularly in circumstances where data is scattered across geographically disperse nodes. However, due to the need for centralized data processing, long delays and high transmission costs accompany when the communication channel capacity between the client and the central server is limited. Furthermore, due to the central edge server's limited computational power and storage resources, it is difficult to enable the construction of complicated models based on massive datasets using centralized edge learning. Therefore, in order to improve the efficiency of centralized edge learning, the sensing, communication, and computation processes need to be jointly designed and the associated resources should be judiciously managed as described in the sequel.

\subsection{Joint Communication and Computation Resource Management in Centralized Edge Learning}
As previously stated, centralized edge learning may introduce a significant delay, which will be catastrophic in latency-sensitive applications such as autonomous driving and VR games. Furthermore, the massive data communication places a significant strain on the backbone network, resulting in significant computation overhead for the central server. Thus, centralized edge learning exists at the crossroads of two domains: communication and computing. The advent of new disciplines brings many interdisciplinary research opportunities, and joint design is required to interweave the two sectors in order to solve the aforementioned challenges to accomplish fast and efficient intelligence acquisition while allocating suitable resources.
The overall goal of traditional communication systems is to maximize throughput. Differently, centralized edge learning systems aim to maximize learning performance. As a result, if the traditional communication system optimization method is still used, optimal learning performance may not be achieved because specific learning factors such as model and data complexity are not taken into account. For example, because the structures of a support vector machine (SVM) and a convolutional neural network (CNN) differ, they can achieve different learning accuracies with the same training data sample size. Furthermore, the communication cost of transferring a data sample in various ML tasks can vary significantly. As a result, new resource allocation algorithms are required for centralized edge learning.

There have been several prior works proposing optimized resource allocation schemes for centralized edge learning in recent years. One efficient design is to schedule the joint communication and computation processes based on the data importance. In \cite{Liu2021TCCN}, the authors proposed a data-importance-aware user scheduling scheme for edge ML systems by considering specific SVM tasks. In this scheme, a data importance measure is proposed to classify the data into different importance levels and then allocate more resources to the data with higher importance. Furthermore, the design based on SVM in \cite{Liu2021TCCN} was then extended to CNN, in which the authors considered the specific retransmission decision problem in order to ensure the quantity and quality of training data in the face of transmission data errors. Unlike the traditional automatic-repeat-request which focuses solely on reliability, the authors selectively retransmitted data samples based on their importance in order to accelerate the convergence speed and accuracy  of the training. 
The other line of research focuses on the learning-centric wireless resource allocation. In \cite{Wang2020TWC}, the authors proposed and validated a nonlinear classification error model for ML tasks. A power allocation scheme centered on learning performance was proposed based on this model. Using the total variational minimum optimization method, the Karush-Kuhn-Tucker (KKT) solution could be obtained. Furthermore, in \cite{Zhou2021TVT}, the authors employed statistical mechanics of learning to forecast the relationship between the learning accuracy of various tasks and the amount of training data. The learning performance is maximized by using differential convex programming (DCP). According to the analysis, the optimal transmission time is inversely proportional to the generalization error. 

\subsection{Mixup Data Augmentation with AirComp }

AirComp is another integrated communication-and-computation design that can be exploited to enhance the efficiency of centralized edge learning, and particularly combined with the mixup data augmentation. Mixup is a well-known data enhancement technology \cite{Zhang2017arXivMIXUP}. Mixup, which in essence, trains neural networks using convex combinations of data and labels, thus regularizing neural networks to make learning simple linear behaviors between training samples easier.


\begin{figure}[t]
	\centering
	\includegraphics[width=0.9\textwidth]{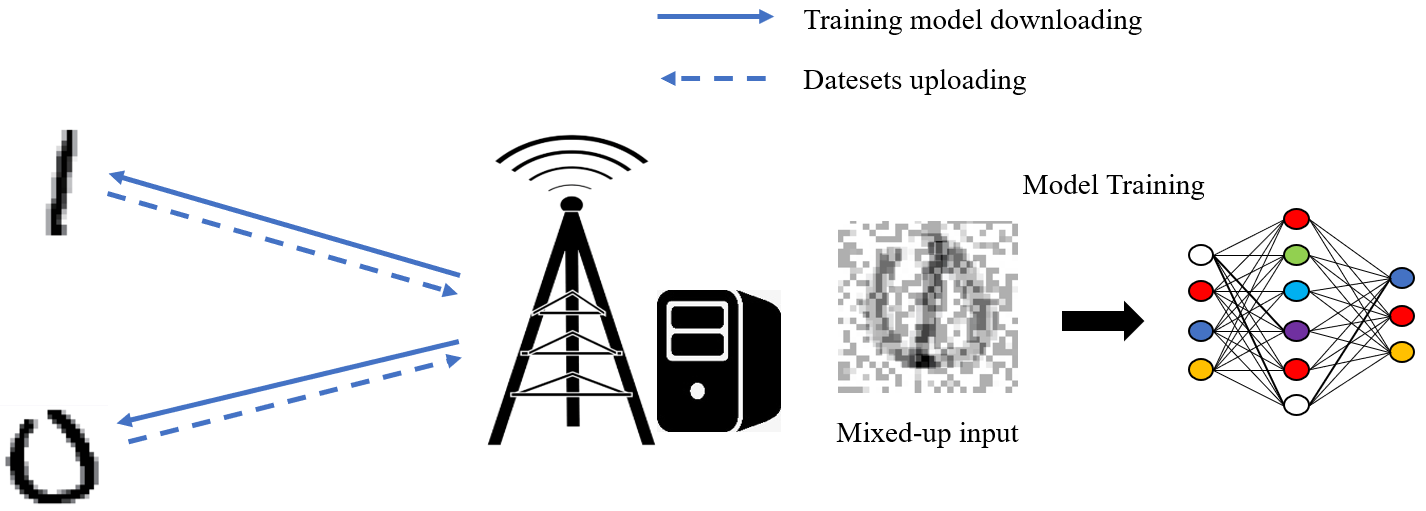}
	\caption{Transmitting procedure in AirMixML.}
	\label{fig:AirMixML}
\end{figure}

As shown in Fig.  \ref{fig:AirMixML}, the authors in \cite{Koda2021GLOBECOM} combined mixup and AirComp, in which a ML framework called over-the-air mixup ML (AirMixML) is proposed to take advantage of the natural distortions and superpositions properties of wireless channels. Multiple users in AirMixML send simulated modulated signals of their private data samples to a central server, which uses the received aggregated noisy samples to train ML models, while protecting the user's privacy. AirMixML was shown to train the model using mixup data enhancement and achieves the same accuracy as the raw data samples. Specifically, AirMixML adjusts the privacy disclosure level of the transmitted signal by controlling the user's transmit power, and the Dirichlet dispersion ratio controls the local power contribution of each worker to the superimposed signal. Given that AirMixML is the first privacy-preserving edge ML framework to use over-the-air signal superposition and additional channel noise without on-device training, there is still much room for improvement before practical application.

\subsection{Centralized Edge Learning with ISAC }

The traditional sensing and communication stages are carried out sequentially. Nevertheless, in ISAC systems, the sensing and communication are implemented via unified wireless signals. This thus motivates many prior works that proposed to accelerate the centralized edge learning by ISAC. In particular, the sensing and communication stages are combined to maximize the use of wireless signals for both dataset generation and transmission. However, ISAC also introduces additional interference between sensing and communication functions to be dealt with.

In order to integrate dataset generation and uploading, in \cite{Zhang2022ICC}, the authors proposed a classification error minimization method for beamforming and time allocation.
Besides, several prior works combined ISAC with data offloading in MEC, which can be utilized in the centralized edge learning with ISAC. In \cite{Ding2022SAC}, the user conducted sensing and communication on the same spectrum using a MIMO array based on the dual-functional radar communication technology. The authors proposed a multi-objective optimization problem for jointly optimizing transmit precoding for sensing, communication and allocating computation resources, by considering both beampattern design and multi-user MIMO radar computation offloading energy consumption. In \cite{Liang2022TWC}, the authors studied the throughput maximization in a multi-user MEC system using a sense-then-offload protocol. Furthermore, in \cite{Qi2021IoT}, the authors investigated a traffic-aware task offloading scheme in a vehicular network and proposed an offloading mechanism based on the sensed environment data. 
In \cite{Roth2022ISJCS}, the author proposed a sensory system for sensing and communication based on analog spike signal processing.

\subsection{Research Opportunities }

Despite the research efforts discussed above for efficient centralized edge learning, there are still many unexplored territories yet, which are discussed as follows.

\begin{itemize}
\item {\bf Secure Data Uploading}: The centralized edge learning architecture may suffer from the \emph{single point of failure} issue, that is, the central server is prone to attack by malicious users who may upload forged or poisoned data to mislead the entire training process. Therefore, how to build a trustworthy mechanism to guard against the potential attack in the data uploading process is a critical issue warrant further study.

\item {\bf Data-importance aware systems}: Due to the limited communication resources in a real communication system, it is often not possible to transmit all of the datasets, so it is necessary to selectively transmit the data in order to train the network more efficiently. Furthermore, the importance of data changes during the transmission process (for example, a certain large category of data is important until it is transmitted, but after a certain amount of data has been transmitted, the same large category of data has little impact on the training), and this aspect has not been considered in previous studies.

\item {\bf Task-oriented ISCC in centralized edge learning}: The various learning tasks in centralized edge learning frequently require the simultaneous support of sensing, communication, and computation functions. As a result, joint resource management for ISCC is required to improve the performance of centralized edge learning. There is still an unexplored territory for task-oriented ISCC targeting edge inference, which motivates the future work's main theme.

\end{itemize}

\section{Federated Edge Learning}\label{sec:FEEL}



FEEL is a machine learning setting where multiple edge devices collaborate in training a shared ML model, and each device’s raw data is stored locally and not exchanged or transferred. From the networking perspective, FEEL can be divided into two classes, including centralized FEEL and decentralized FELL. Centralized FEEL is the most popular FEEL architecture, in which an edge server coordinates the ML model training among the edge devices. Unlike centralized FEEL, decentralized FELL is a network topology without any central server to coordinate the training process, in which all edge devices are connected in a peer-to-peer manner to perform the model training. For FEEL, the sensing, communication and computation are three important functions that must be involved in completing a learning task using FEEL. To begin with, the edge devices must sense the environment (e.g., by using the equipped radio sensors) in order to obtain data for training ML models; Second, the edge devices compute model updates using local computing power; Finally, the edge devices upload local model updates to the edge server via the uplink channel, and the edge server broadcasts the global model to each edge device via the downlink channel. Wireless communication, sensing, and computation all have different effects on federal edge learning, however, in edge networks where resources (e.g., radio bandwidth and edge device energy supply) are typically limited, and the three are frequently coupled and constrained by each other, thus making their joint design necessary.

\subsection{Joint Communication and Computation Resource Management in FEEL}

In FEEL, the heterogeneity of edge devices in terms of radio sources, channel status, and computational capabilities is a common issue that has a direct impact on the ultimate learning performance such as accuracy and latency. This thus calls for joint communication and computation resource allocation for FEEL performance optimization.  Specifically, the mentioned device heterogeneity essentially leads to 
distinct uploading time between different devices and edge server. As a result, if the edge server uses synchronized aggregation of the model updates of the edge devices, the edge device with the longest delay dominates the communication time of a single round. 
To address this issue, the authors in \cite{Luo2022Infocom} investigates the optimal scheduling scheme for edge devices to minimize the training time of federal edge learning, but the communication resource optimization is not involved. 
In \cite{Chen2022IoTJ-RA}, the heterogeneous channel conditions of edge devices are also considered, in which the objective is to maximize the number of participating edge devices, by optimizing the scheduling scheme of edge devices and the allocation of communication resources including transmit power and bandwidth. 
In addition to the heterogeneity in channel conditions, the edge devices tend to be heterogeneous in computing capabilities as well. 
For instance, both \cite{Chen2020TWC-joint} and \cite{Xu2021TWC-scheduling-RA} comprehensively consider the heterogeneity of edge devices in terms of channel conditions and computing capabilities, and study the optimal scheduling scheme for edge devices and the optimal communication resource allocation.
The difference between the \cite{Chen2020TWC-joint} and \cite{Xu2021TWC-scheduling-RA} is that the considered problem in \cite{Chen2020TWC-joint} only focuses on a single communication round and each communication round is treated equally. 
In contrast, \cite{Xu2021TWC-scheduling-RA} explicitly takes the importance of different communication rounds into consideration and investigates the bandwidth allocation and edge device scheduling problems under long-term energy constraints. Moreover, imperfect wireless channel conditions are also investigated in \cite{Xu2021TWC-scheduling-RA}.

On the other hand, as edge devices need to utilize local computing resources for model updates. Research on computing resource management mainly focus on two directions: optimization of the CPU/GPU frequency of edge devices \cite{Nguyen2020FEEL-IoT,Dinh2021ACM-FL-RA} and optimization of the batch size used in model updates (e.g., stochastic gradient descent) \cite{Ma2021TMC-Batch,Battiloro2021Lyapunov}. Since edge devices are usually heterogeneous in terms of computing power, the time for completing training and the energy consumed by federal edge learning can be largely wasteful if the computing power of different edge devices is not reasonably tuned. Both the literature \cite{Nguyen2020FEEL-IoT} and \cite{Dinh2021ACM-FL-RA} consider the total energy consumption of the system and the required training time simultaneously as minimization objectives to optimize the CPU frequency of different edge devices, as well as system variables such as communication resources and device scheduling. For the case where the edge devices cannot effectively adjust the computing frequency, \cite{Ma2021TMC-Batch} optimizes the batch size of different devices to align the communication delay between different edge devices and the edge server, thus reducing the training time for federal edge learning. Unlike \cite{Ma2021TMC-Batch}, which only considers the optimization within a single communication round, literature \cite{Battiloro2021Lyapunov} focuses on the whole federated edge learning training process, where the authors consider a long-time dynamic resource optimization problem, and a scheme based on Lyapunov optimization is proposed to jointly optimize the computing frequency and the batch size of each edge device in different communication rounds. 

Besides the heterogeneity of communication and computation, the data heterogeneity is also typical in federated learning, as the data at each edge device is likely to be personalized towards the specific device. A few works tended to manage the communication and computation resources in FEEL, by considering the effects of data heterogeneity. For example, in \cite{Luo2022Infocom_noniid}, the optimal client sampling strategy that tackles both system and data heterogeneity is designed to minimize the training time with convergence guarantees in a FEEL system with resource-constrained devices. The work in \cite{Liu2022FITEE} considered quantized FEEL with data heterogeneity, and jointly optimized the quantization level and the bandwidth allocation to minimize the training time.

\subsection{ Over-the-Air Federated Edge Learning (Air-FEEL)}

\begin{figure}[t]
	\centering
	\includegraphics[width=0.75\textwidth]{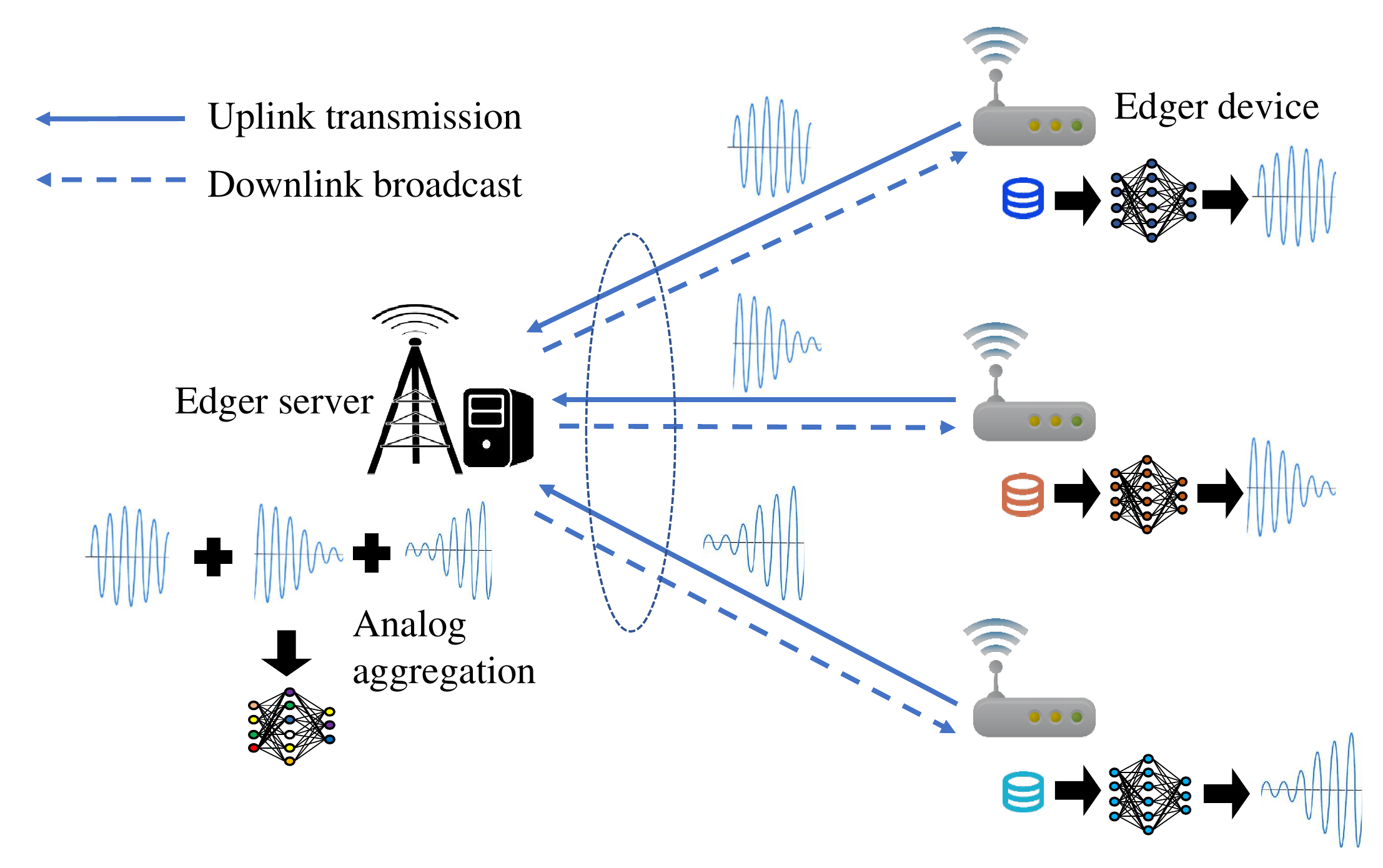}
	\caption{Illustration of the Air-FEEL system.}
	\label{fig:Air-FEEL}
\end{figure}

Air-FEEL has emerged as a promising solution for edge AI \cite{Cao2022AircompOverview}. As shown in Figure \ref{fig:Air-FEEL}, over-the-air model/gradient aggregation is used in Air-FEEL to improve learning efficiency of FEEL. It is shown in \cite{Zhu2019TWC_broadband} that, compared with the conventional orthogonal multi-access, Air-FEEL can reduce the communication latency by a factor approximately equal to the number of devices without significant loss of the learning accuracy. Various research efforts have focused on different directions in Air-FEEL, such as resource management, device scheduling, and privacy preserving schemes. 

In Air-FEEL, edge devices can control their transmit power adaptively to reduce aggregation error for model/gradient aggregation, and thus improve learning accuracy or convergence rate. The authors in \cite{Zhang2021TWCgradient} developed the optimized power control to minimize the AirComp MSE, which has a threshold-based structure depending on the channel conditions at different edge devices. Rather than minimizing the aggregation MSE in each round independently, the authors in \cite{Cao2021arXiv_power,Cao2022JSAC-aircomp-average} optimized power allocation across multiple global rounds to maximize the convergence rate in terms of the loss optimality gap. Device scheduling is another efficient technique for improving Air-FEEL performance via addressing resource heterogeneity by dropping edge devices with poor communication and computation conditions. 
The work \cite{Yang2020TWC} presented a joint design of device scheduling and receive beamforming in which the edge devices with weak signal strengths after receive beamforming were dropped from the training process. Furthermore, the work \cite{Sun2022JSACaircompDynamic} investigated device scheduling by taking into account their diverse energy constraints and computation capabilities, and an energy-aware dynamic device scheduling algorithm based on Lyapunov optimization was proposed.
Utilizing Air-FEEL offers an additional advantage in improving data privacy, in addition to the benefit of reducing multiple-access latency. The local training data can still be inferred from the local model updates even with powerful model inversion techniques \cite{Haque2020Nature}. As a fix, Air-FEEL limits eavesdroppers' access to the aggregated updates, hiding each individual local update in the sea of others. A further free mask that can be used to safeguard the privacy of the data is the random disturbance that the channel noise imposes on the aggregated updates \cite{Liu2020JSAC_privacy}. A more comprehensive overview on Air-FEEL can be referred to \cite{Cao2022AircompOverview}.

Although AirComp is beneficial for model aggregation in Air-FEEL due to the inherent superposition property of wireless channels, Air-FEEL also suffers from the straggler issue in which the device with the weakest
channel acts as a bottleneck of the model aggregation performance. To address this issue, the work in \cite{Liu2021TWC_FEEL_RIS} leveraged the RIS technology in Air-FEEL, and a unified communication-learning optimization problem is solved to jointly optimize device selection, over-the-air transceiver design, and RIS configuration.
The aforementioned studies all focused on the centralized FEEL, in which a central edge server is required to orchestrate the training process. The work in \cite{Shi2021ISIT_DFL,Ozfatura2020GC_DFL} considered decentralized FEEL in the scenario where an edge server is not available or reliable, in which the authors considered the precoding and decoding strategies for device-to-device communication-enabled model/gradient aggregation and proposed an AirComp-based decentralized stochastic gradient decent with gradient tracking and variance reduction algorithm to reach the consensus. 

\subsection{Federated Edge Learning with ISAC }

\begin{figure}[t]
	\centering
	\includegraphics[width=0.75\textwidth]{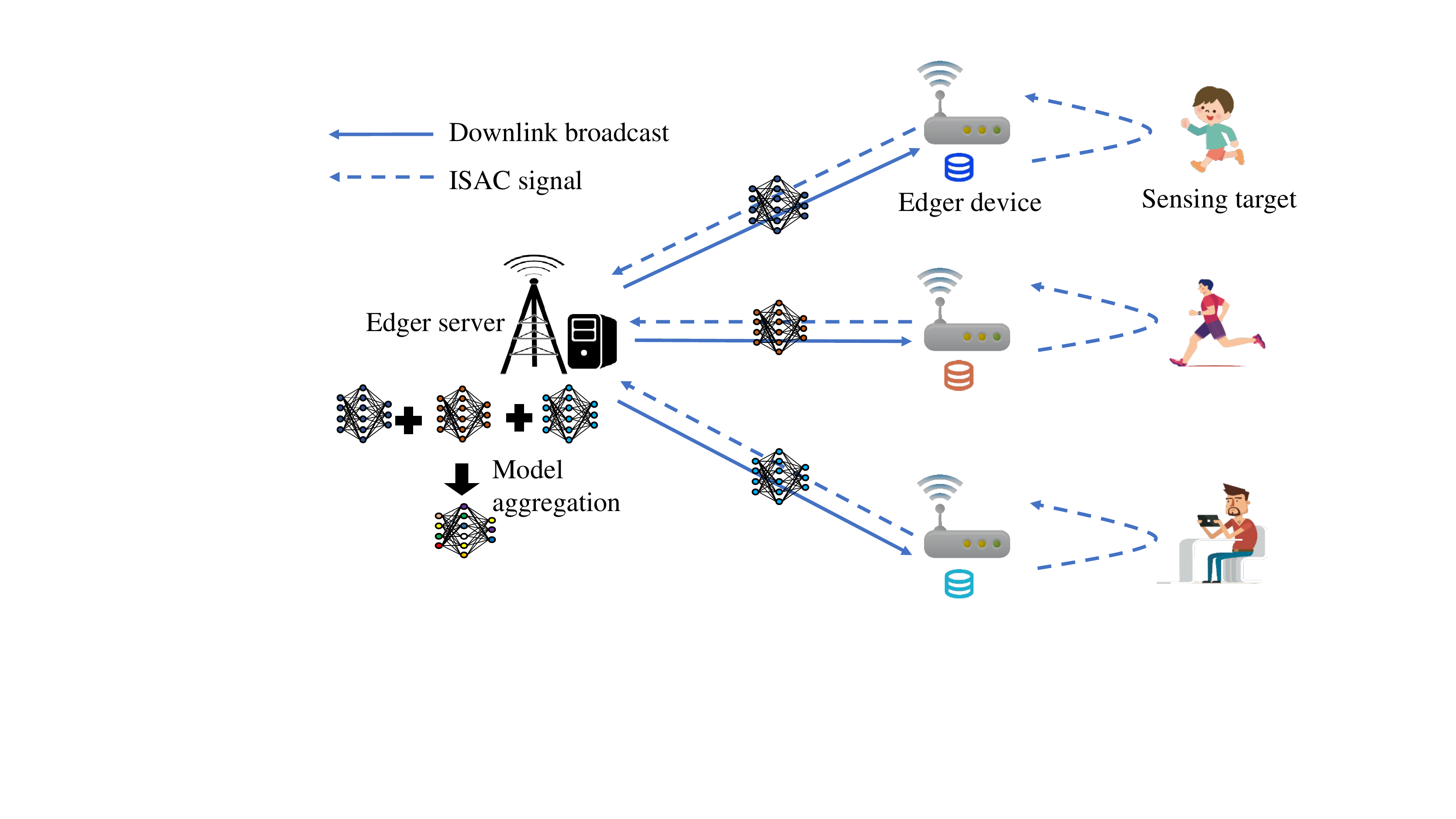}
	\caption{Illustration of the FEEL system with ISAC.}
	\label{fig:FL-ISAC}
\end{figure}

Federated edge learning also be exploited to training AI models in wireless sensing networks, and in this case, ISAC can be integrated to enhance the training and sensing performance, as shown in Figure \ref{fig:FL-ISAC}.  However, it is still an open problem to design edge learning systems with sensing and communication coexistence \cite{Liu2021ISAC-survey}.

In ISAC, sensing and communication can be integrated in three different levels. At the first level, the sensing and communication signals may occupy orthogonal time-frequency resources, which do not interfere with each other functionally but compete for time-frequency resources, so the reasonable resource allocation between sensing and communication is a key issue when communication and sensing orthogonally coexist. For example, in \cite{Li2021sensing-model-journal}, the sensing and communication work in a time-division manner, so that sensing and communication jointly compete for time resources. Based on this, a mathematical relationship between sensing time and learning performance is experimentally fitted and the trade-off between learning performance and communication rate is analyzed in \cite{Li2021sensing-model-journal}.
The work \cite{Liu2022arXivISCC} considered horizontal FEEL with ISAC and jointly optimizes the sensing, communication and computation resources, where sensing and communication occupy the same frequency band and different time durations, similarly in \cite{Li2021sensing-model-journal}.
At the second integration level, sensing and communication may work on the same time-frequency resources but use separated signal waveforms, and they will interfere with each other, so how to manage interference becomes especially important.
In \cite{Zhang2021ISAC-edge}, the relationship between learning performance and sensing/communication resources is obtained by considering sensing and communication working on the same time-frequency resources and linking the learning performance to the quantity of sensed samples. Since sensing and communication may interfere with each other spatially, in \cite{Zhang2021ISAC-edge}, the beam directions of the perception and communication signals are optimized with the goal of maximizing the learning performance. 
In the third level,  sensing and communication functions are simultaneously implemented using a shared signal waveform, which can effectively improve the system spectrum efficiency, hardware efficiency, and information processing efficiency \cite{Liu2021ISAC-survey,Cui2021Netw-ISAC-survey,Liu2021ISAC-survey-fundamental}.
The literature\cite{Li2022arXiv-ISAC-aircomp} is the first to combine level-three ISAC with over-the-air analog aggregation technology and jointly designed beamforming for both sensing and communication signals, laying the foundation for the subsequent application of ISAC to federated edge learning. Unleashing the full potential of FEEL by combining with ISAC has very significant application prospects. 
In order to save frequency resources, the same ISAC signal is adopted for both sensing and communication in \cite{Liu2022CLfediSAC}, and a cooperative sensing scheme based on vertical FEEL is proposed to enhance the sensing performance.
\color{black}

\subsection{Research Opportunities}
Despite the research efforts discussed above for efficient FEEL, here lists some unexplored problems and challenges to motivate future works.

\begin{itemize}
	\item {\bf Multi-modal data sources}: In real-world sensing systems, the edge devices involved in training may have different types of sensors, such as radio sensors or cameras, and the sensed data may have different modalities \cite{Guo2022SPL-multi-modal}. In such cases, a FEEL system with ISCC for multimodal data needs to be designed and optimized.
	
	\item{\bf Dynamic sensing environments}: Most of the current works consider static sensing environments, but in real scenarios the sensing environment may keep changing over time and the distribution of the sensed data samples will no longer be stationary. How to design a federated edge learning ISCC system for dynamic sensing environments is also a topic worthy of in-depth study.
	
	\item{\bf Task-oriented ISCC in FEEL}: Few current studies on FEEL have considered specific sensing process, mostly focusing on resource optimization for communication and computation. Various types of learning tasks in FEEL often require the support of sensing, communication, and computation functions at the same time, resulting in a variety of complex relationships among the sensing, communication, and computation modules, such as coupling, collaboration, and even competition. Therefore, in order to improve the upper limit of the performance of FEEL, joint resource management of sensing, communication, and computation is required.
\end{itemize}

\section{Edge Inference}\label{sec:inference}

Apart from the training phase discussed above, edge inference is another important aspect for supporting the successful implementation of AI technologies at wireless edge \cite{9562559}. Specifically, for edge inference, a well-trained ML model needs to be deployed at network edge to run AI tasks in real time (such as classification, recommendation, and regression), which is beneficial to computation/storage/power-limited edge devices and delay-sensitive AI tasks \cite{WXu}. Thus, edge inference has become an important technique to enable various AI applications, such as auto-driving and smart cities in 6G networks. For example, in auto-driving, the vehicles need to detect obstacles to avoid accidents under stringent latency constraints. To guarantee high detection accuracy, more sophisticated DNN architectures are preferable for the detection, say, ResNet-50 \cite{Resnet}. However, ResNet-50 contains 50 convolutional layers, and demands nearly 100 megabytes of memory for storage. On one hand, it is non-trivial to deploy such complicated DNN on edge devices with limited computation and storage capacity. In this case, it merely on the cloud server as this will induce intolerable delay and may increase the risk of accidents. To deal with such dilemma, edge inference provides a promising solution with better trade-off among computation power, storage, and communication latency. 

There are three different methods to implement edge inference, namely on-device inference \cite{SFYilmaz}, on-server inference \cite{YangK2020,SHua2021}, and split inference \cite{M.Jankowski,M.Jankowski2020,JShao,FPezone,JShao2021,HXie,XHuang,XTang,JShao2020,ZLiu,QLan,Wen,lee2022,Wen2022}. For on-device inference, the computation is accomplished merely on edge devices, which is non-trivial for recent increasingly complex AI models and computation/storage/power-limited edge devices. To tackle such issue, on-server inference is designed. However, on-server inference suffers from the communication bottleneck due to the potential high-volume data transmission over the band-limited wireless channels in the presence of uncertain channel fading, under the stringent low-latency constraint. Also, the computation resources at edge servers may still fall short when running some large-scale AI tasks. Nevertheless, potential information leakage during data uploading from the edge devices to the edge server may lead to privacy issues in edge inference. To tackle these problems, split inference is proposed jointly considering techniques such as joint source and channel coding (JSCC)\cite{M.Jankowski,M.Jankowski2020,NTishby,JShao,FPezone,JShao2021,HXie}, joint communication and computation resource management design \cite{XHuang,XTang,JShao2020,ZLiu,QLan}, and AirComp \cite{Wen,lee2022}. Furthermore, as a recently proposed technique, ISAC has drawn increasing research interests \cite{Liu2021ISAC-survey}. Intuitively, ISAC can further reduce the latency of edge inference due to the integrated data sensing and uploading process \cite{Wen2022}. The joint management of sensing, communication, and computation resources in this case is more difficult. In the following, we discuss the above techniques in detail.

\subsection{JSCC in Edge Inference}
Generally, when uploading features from the devices to the serves to perform inference, the fluctuating wireless channels may introduce lossy transmission, which calls for proper design of JSCC for efficient feature transmission. Moreover, in some sense, JSCC can be viewed as a novel design principle bringing both communication and computation into consideration. With the recent development of DL, deep JSCC has been widely investigated to alleviate excessive signaling overhead as well as improve the robustness to channel distortion. For example, for classification tasks, the authors in \cite{M.Jankowski} and \cite{M.Jankowski2020} proposed a retrieval-oriented wireless image transmission framework to maximize the classification accuracy, where the JSCC framework is trained by the cross-entropy between the predictions and the ground-truth labels. Moreover, information bottleneck (IB) \cite{NTishby} was proposed to extract minimum features to fulfill certain tasks sufficiently. Under the guidance of IB, an image classification task was considered in \cite{JShao} by designing a framework of task-oriented JSCC. By combining IB with stochastic optimization, the same task was considered in \cite{FPezone} to minimize energy consumption as well as service latency simultaneously. Considering edge inference with multiple edge devices, a task-oriented JSCC framework was designed in \cite{JShao2021}, where a group of edge nodes perform the classification task coordinated by an edge server. Furthermore, some initial exploration has been made in \cite{HXie} to deal with the multi-modal data, where a task-oriented semantic communication scheme was proposed and the cross-entropy objective with multiuser multi-modal data fusion was considered.

\subsection{Joint Communication and Computation Resource Management in Edge Inference }

Edge inference typically involves local feature extraction and uploading to the edge server for further processing, which may encounter both communication and computation bottleneck. Specifically, on one hand, transmitting data through wireless links naturally suffers from channel impairment, especially when the AI services have stringent low-latency requirement. On the other hand, processing data at devices and servers incurs computational delay, especially when a large number of devices require for AI services simultaneously. Furthermore, there always exists a trade-off between inference accuracy and the computation-communication capability of the system in edge inference.

To address the above issues, joint management of communication and computation resources is considered in recent works. For instance, in \cite{XHuang}, the optimal control of inference accuracy and transmission cost was modeled as a Markov decision process (MDP), and their trade-off is balanced via dynamically selecting the optimal compression ratio with hard deadline requirements. By further considering proper split of AI models in multi-user edge inference systems, the authors in \cite{XTang} jointly designed the model split point selection and computational resource allocation to minimize the maximum inference latency. Also, the authors in \cite{JShao2020} studied the trade-off between the computational cost of the on-device model and the communication overhead of uploading the extracted features to the edge server. Then they propose a three-step framework for inference, which contains model split point selection, communication-aware model compression, and task-oriented encoding mechanism for the extracted features. Due to the heavy computational burden being offloaded on the server, it is also urgent to study the computation and inference accuracy trade-off at the server side. In \cite{ZLiu}, the authors considered the early exiting technique, which allows a task exit from certain layers of a DNN without traversing the whole network. In such way, the joint management of communication and computation resources could reduce the inference latency under various accuracy requirements. Moreover, a progressive feature transmission protocol was proposed in \cite{QLan}, which contains importance-aware feature selection and transmission-termination control. In such a protocol, the devices transmit the extracted features progressively according to their importance, and once the inference accuracy requirement is obtained, the transmission will stop. With such design, the trade-off between inference accuracy and communication-computation latency can be well balanced. 
Moreover, RIS  has recently emerged as a potential solution to provide a cost-effective way for enhancing the performance of edge inference. For example, a RIS-aided green edge inference system was considered in \cite{9352968}, where the set of tasks performed by each BS, uplink/downlink beamforming vectors of BSs, transmit power of edge devices, and uplink/downlink
phase-shift matrices at the RIS were jointly designed to  minimize the overall network power consumption.
\subsection{Over-the-air Edge Inference }
\begin{figure}[t]
\centering
    \includegraphics[width=12cm]{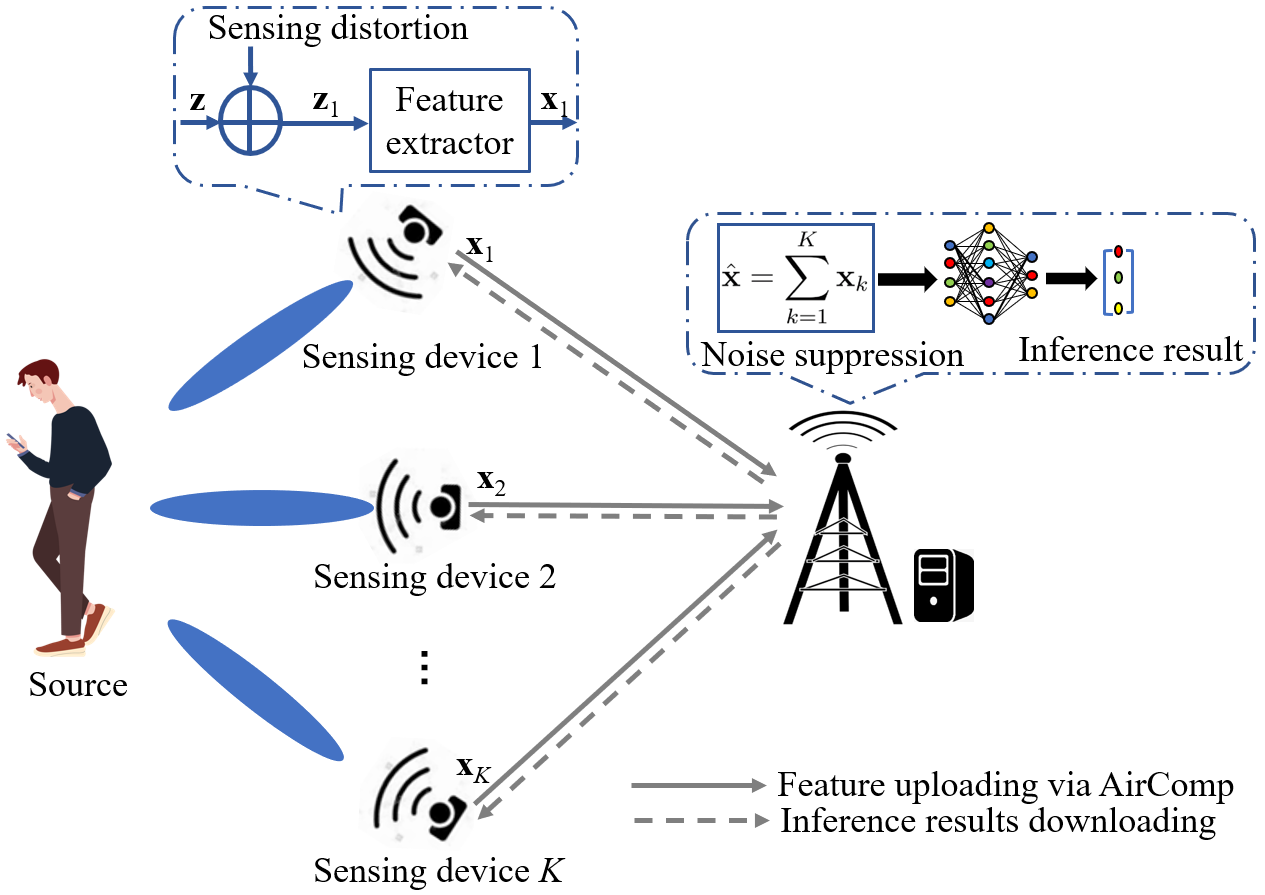}
\caption{Illustration of AirComp-based edge inference systems.}
\label{Aircomp_inference}
\end{figure}

AirComp is also appealing for low-latency edge inference by seamlessly integrating communication and computation, as shown in Fig. \ref{Aircomp_inference}. The research on over-the-air edge inference is still in its early stage.

An initial study of AirComp-based multi-device edge inference system was made in \cite{Wen}, where AirComp is utilized to aggregate multiple noisy feature observations of a common source to average out the feature noise for boosting the inference accuracy. The authors first characterize the influence of sensing and channel noise on inference accuracy by deriving a tractable surrogate performance metric called discriminant gain. Then the authors maximize the inference accuracy by jointly optimizing the transmit precoding and receive beamforming. Besides the significantly enhanced spectrum efficiency, AirComp has additional benefit in privacy preservation. To exploit such property, the authors in \cite{SFYilmaz} considered an ensemble inference framework, where each device needs to transmit their predictions to the server for further fusion to obtain final results. Apart from maximizing inference accuracy, the authors also consider maximizing the privacy of the on-device models. To this end, AirComp is exploited for privacy-enhanced outcome fusion as each individual predictive outcome is hidden in the crowd. Specifically, the authors introduce different ensemble methods, such as belief summation and majority voting, and provide privacy analysis for these AirComp-based fusion schemes. Finally, numerical results provided in \cite{SFYilmaz} have shown that the proposed AirComp-based solution significantly outperforms other orthogonal transmission schemes in terms of the required communication overhead under the same target privacy guarantee. Moreover, the authors in \cite{lee2022} also considered the privacy issues in edge inference. Specifically, they consider the distributed inference of graph neural networks (GNNs). To deal with the possible privacy leakage problem arising from the devices exchanging information with neighbors during decentralized inference, the authors first characterize the privacy performance of the considered decentralized inference system. Then they design privacy-preserving signals and the corresponding training algorithms in combination with AirComp to further boost the privacy of the considered system.

\subsection{Co-inference with ISAC  }
\begin{figure}[t]
\centering
    \includegraphics[width=12.5cm]{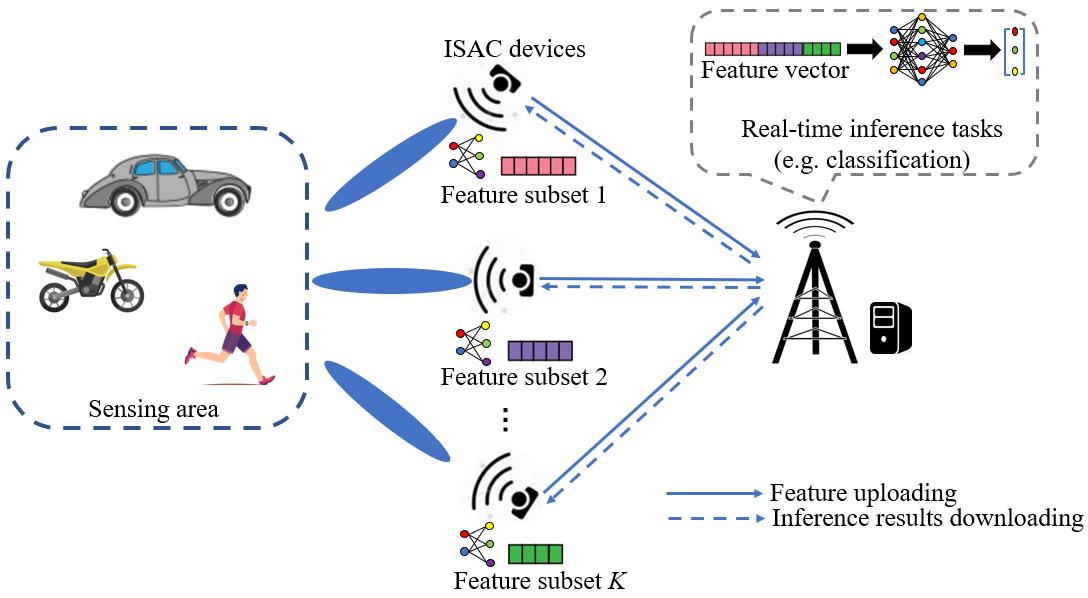}
\caption{Edge inference systems with multi-device sensing.}
\label{ISAC_inference}
\end{figure}
In future wireless networks, to support environment-aware intelligent applications,
it is desirable to process and upload the collected data from sensing devices for inference, where sensing,
communication, and computation are naturally coupled and need to be jointly designed. However, integrating ISAC with edge inference introduces several issues. First, it is non-trivial to characterize the inference performance in ISAC-enabled edge inference. Second, how to jointly design the resources for sensing, communication, and computation to proper balance the trade-off among them is challenging. To deal with the above issues, the authors in \cite{Wen2022} studied a task-oriented ISCC-based edge inference system as shown in Fig. \ref{ISAC_inference}, where multiple ISAC devices collect sensing data, and then upload the quantized features to the server for classification. The authors analyze inference performance in such ISCC-based system via deriving the tractable discriminant gain, based on which, the allocation of sensing, transmit power, communication time, and quantization bits are jointly designed for the successful completion of the subsequent classification task. Finally, some interesting design insights for balancing the trade-off between sensing, communication and computation were crystalized in \cite{Wen2022}: the sensing power and quantization bits should be enlarged as the number of classes increases in the classification task, otherwise, more communication power should be allocated if the channel conditions of the devices are poor.

\subsection{Research Opportunities  }
Despite the research efforts discussed above for efficient edge inference, there are still many open problems and challenges unexplored yet.

\begin{itemize}

    \item {\bf Fundamental limits of JSCC:} Similar as Shannon’s information theory, the fundamental limits of semantic based JSCC transmission for inference needs to be characterized. Moreover, it is also interesting to explore the number of optimized symbols for the successful completion of certain tasks via JSCC transmission, which can  balance the delay and accuracy trade-off in JSCC-based edge inference.

 \item {\bf Case with  multiple devices and/or multiple servers:} For edge inference systems with  multiple devices and/or multiple servers, the device selection and server coordination need to be further considered. On one hand, the selection of devices for inference need to account for the trade-off between delay and accuracy. On the other hand, multi-server system requires flexible model deployment for large-scale AI tasks, such as heterogeneous tasks with different models requirements.


\item {\bf Task-oriented ISCC for edge inference:} In edge inference, sensing, communication, and computation may compete for resources (such as radio and hardware). How to depict the relationship between the inference performance and all the three mentioned processes  is quite challenging, thus yielding the non-trivial problem of joint management of sensing, communication, and computation resources. Although \cite{Wen2022} did an initial study in this direction, there still remain many uncharted issues warrant further investigation. Moreover, to fully exploit edge intelligence, multi-modality sensors (such as laser radar, millimeter-wave radar, and cameras) may be deployed at wireless edge. How to process the acquired multi-modal sensing data for efficient inference is also an interesting direction to pursue.
\end{itemize}

\section{Concluding Remarks}\label{sec:conclusion}
In the upcoming 6G era, we shall witness the paradigm shift in network functionality from connecting people and things to connecting intelligence, driving the advancement of IoT in 5G towards AIoT. As a consequence, with the assist of 6G networks, AI is expected to spread from the cloud to the network edge so as to deliver pervasive AI services. However, classic design principle of separating sensing, communication, and computation cannot meet the stringent demand in terms of latency, reliability, and capacity. 

To tackle this issue, this article presented a timely literature survey on the tasked-oriented ISCC for edge intelligence. First, we introduced the motivations and basic principles on ISCC. Then, we introduced representative works on three different scenarios, i.e., centralized edge learning, federated edge learning, and edge inference, respectively, by focusing on the joint communication and computation resource management, AirComp, and ISAC, in each scenario. Finally, interesting research directions were presented to motivate future works. It is our hope that this article can provide new insights on this interesting research topic, and motivate more interdisciplinary research from the communities of wireless sensing, wireless communications, machine learning, and computing.






\bibliographystyle{IEEEbib}
\bibliography{references_Mingzhe}



\end{document}